\documentclass[dvips]{article}
\usepackage{icrctc07}

\title{The ANTARES neutrino telescope: a status report}
\shorttitle{Antares first results}
\authors{A. Kouchner$^{1}$ on behalf of the Antares collaboration}
\shortauthors{A. Kouchner et al}
\afiliations{$^1$Laboratoire AstroParticle and Cosmology \\ 10, rue A. Domon et L. Duquet,
75205 Paris Cedex 13, France.}
\email{kouchner@apc.univ-paris7.fr}

\abstract{ANTARES is a large volume neutrino telescope currently under construction off La Seyne-sur-mer, France, at 2475m depth. Neutrino telescopes aim at detecting neutrinos as a new probe for a sky study at energies greater than 1 TeV. The detection principle relies on the observation, using photomultipliers, of the Cherenkov light emitted by charged leptons induced by neutrino interactions in the surrounding detector medium. Since late January 2007, the ANTARES detector consists of 5 lines, comprising 75 optical detectors each, connected to the shore via a 40 km long undersea cable. The data from these lines not only allow an extensive study of the detector properties but also the reconstruction of downward going cosmic ray muons and the search for the first upward going neutrino induced muons.The operation of these lines follows on from that of the ANTARES instrumentation line, which has provided data for more than a year on the detector stability and the environmental conditions. The full 12 line detector is planned to be fully operational early 2008.}

\begin{document}
\maketitle

\section{Scientific motivations}

One of the major aims of neutrino astronomy is to contribute solving the fundamental question 
of the origin of high energy cosmic rays (HECR).
Neutrinos can indeed escape from the 
core of the sources and travel with the speed of light through magnetic 
fields and matter without being deflected or absorbed. 
Therefore they can deliver direct information about the processes taking 
place in the core of the production sites and reveal the existence of 
undetected sources. At high energies, neutrinos are 
unmatched in their capabilities to probe the Universe. 

High energy neutrinos are produced in a beam dump
scenario in dense matter via pion decay, when the accelerated protons 
interact with ambient matter or dense photon fields:
{\small
\[
  p + A/\gamma \rightarrow
\begin{array}[t]{l}
 \pi^0 \\
 \downarrow \\
 \gamma \gamma
\end{array} +
\begin{array}[t]{l}
 \pi^\pm \\
 \downarrow \\
 \mu^\pm + \nu_\mu ~(\overline{\nu}_\mu) \\
 \downarrow \\
 e^\pm +\nu_e ~(\overline{\nu}_e) + \overline{\nu}_\mu ~(\nu_\mu)
\end{array} + N +...
\]
}
Good candidates for high energy neutrino production are active galactic
nuclei (AGN) where the accretion of matter by a supermassive 
black hole may lead to relativistic ejecta~\cite{agn_jet}. Other potential sources of extra-galactic 
high energy neutrinos are transient sources like
gamma ray bursters (GRB). As many models~\cite{grb_review} for GRBs involves the collapsing of a 
star, acceleration of hadrons follows naturally. The diffuse flux of high 
energy neutrinos from GRBs is lower than the one expected from 
AGNs, but the background can be dramatically reduced by 
requiring a spatial and temporal coincidence 
with the short electromagnetic bursts detected by a satellite.

High Energy activity from our Galaxy has also
been reported by ground based gamma-ray telescopes. Many astrophysical sources~\cite{galactic_sources} are 
candidates to accelerate hadrons and subsequently produce neutrinos. Such sources could only be observed 
by a northern neutrino telescope like Antares.

Neutrino telescopes are also sensitive to signals due to the 
annihilation of neutralinos, gravitationally trapped inside the core of massive objects like the Sun, 
the Earth or the Galactic centre~\cite{dm}. 

Finally, deep-sea neutrino telescopes enable researches in the fields of marine biology, 
oceanography and seismology. 

\section{Detection principle}

The neutrino's advantage, the weak coupling 
to matter, is at the same time a big disadvantage.
Huge volumes need to be monitored to compensate for the feeble signal
expected from the cosmic neutrino sources. In this context, the water Cherenkov technique offers both a cheap 
and reliable option.

The detection principle relies on the observation, using a 3 dimensional array of photodetectors, of the 
Cherenkov light emitted, in a transparent medium, by charged leptons induced by charged-current neutrino 
interactions in the surrounding detector medium. 

Thanks to the large muon pathlength, the effective detection volume in the muon channel is substantially 
higher than for other neutrino flavours. The higher the neutrino energy the smaller the deviation between the muon and the neutrino
(typically $\Delta \theta \simeq \frac{0.7^{\rm o}}{(E_\nu {\rm(TeV)})^{0.6}}$), 
thus enabling to point back to the source with a precision close to the one achieved by gamma-ray telescopes. 
Muon trajectories are reconstructed using the time and amplitude from the photodetector signals.

The energy of the event is estimated thanks to the energy deposited in the detector. 
Monte Carlo simulations for sea water predict a muon energy estimation by a factor of 2-3.

Cosmic particles penetrating the atmosphere undergo a cascade of many secondary particles.
Among them, high energy muons can reach the detector and constitute 
a very intense source of background. To suppress this background the detector concentrates 
on upward detection. As a result, the field of view is restricted to 
one half of the celestial sky ($2\pi$~sr). Severe quality cuts criteria are then applied to 
the reconstruction to remove remaining mis-reconstructed muons.
Atmospheric neutrinos produced in the atmospheric cascades can travel through the Earth
 and interact in the detector vicinity. To some extent this background is irreducible. 
Fortunately, the atmospheric neutrino flux shows a dependency upon energy $dN/dE\propto E^{-3.7}$ 
while cosmic neutrinos are expected to exhibit a flux dependency $dN/dE\propto E^{-2}$. 
An excess of events above a certain energy can therefore be attributed to extraterrestrial 
neutrinos.

\section{The Antares Detector}

\begin{figure}[t]
\begin{center}
\includegraphics [width=0.48\textwidth]{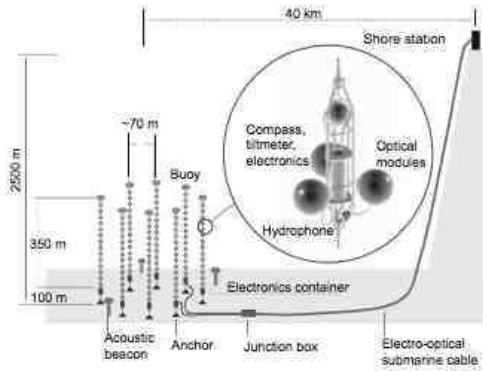}
\end{center}
\caption{Schematic layout of the future Antares detector. The full detector will 
consist of 12 lines connected to a junction box (deployed in December 2002) and operated 
from shore in remote mode through an electro-optical cable. 
}\label{detector}
\end{figure}

Antares is a large European collaboration\footnote{for a complete list of the antares members 
see http://antares.in2p3.fr} currently deploying a 2475~m depth detector $40$~km off 
La-Seyne-sur-Mer (Var, French Riviera) at a location $42^{\rm o}50'N,6^{\rm o}10'E$. 
The site benefits from the close infrastructures of the French sea science institute IFREMER. 
The sea water properties have been extensively studied revealing low light scattering, mainly 
forward~\cite{light} and an average optical background (induced by bacteria and $^{40}K$ decays) 
of 70 kHz per detection channel.

The final detector will consist of an array of 12 flexible individual mooring lines separated 
from each other on the sea bed by 60-80~m. Figure~\ref{detector} shows a sketch of the detector.
The lines are weighted to the sea bed and held nearly 
vertical by syntactic-foam buoys. Each line will be equipped with 75 photomultipliers~\cite{om} 
housed in glass spheres, referred to as optical modules (OM).  The OMs are inclined by $45^{\rm o}$ 
with respect to the vertical axis to ensure maximum sensitivity to upward moving Cherenkov light fronts. 
Expected performances, in particular in the frame of point source searches are described in~\cite{juanan}.

The default readout mode~\cite{daq} of the detector is the transmission of the time and amplitude
 of any light signal above a threshold corresponding to 1/3 of a photo-electron for each OM. Time 
measurements are relative to a master reference clock signal distributed to each storey from 
shore via an electro-optical cable. 
The grouping of three optical modules in a storey allows local coincidences to be made to 
eventually reduce the readout rate. In addition the front end electronics~\cite{ars} allows a more 
detailed readout of the light signal than the standard time and amplitude mode. With this detailed 
readout it is possible to sample (up to 1~GHz) the full waveform of the signal with 128 channels, 
enabling special calibration studies of the electronics.

\section{First results from deep-sea}

\begin{figure}[t]
\begin{center}
\includegraphics [width=0.35\textwidth]{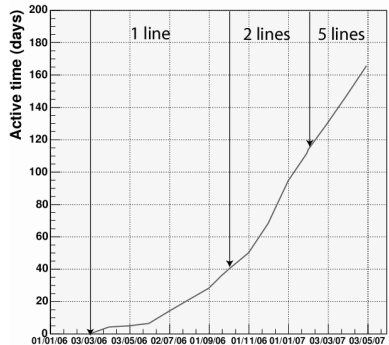}
\end{center}
\caption{Integrated number of effective days of data taking since March 2006
taking into account all losses.}
\label{dataEff}
\end{figure}

A mini-instrumented line equipped with 3 OMs (MILOM) and mainly dedicated to study environmental parameters
(sea current, salinity, pressure, temperature...) has been in operation since spring 2005.
The results of this line are presented in details in~\cite{milom}. 
Since the end of January 2007, the detector consists of 5 operating detection lines. 
At this stage, Antares is the largest 
neutrino telescope ever built in the northen hemisphere. 
Data with 2 lines have been taken since October 2006 and with one line since March 2006. Figure~\ref{dataEff} 
gives an indication of the data taking efficiency since the connection of the first line, which has been 
continuously improving. In spring of 2007 two further lines were immersed and two more lines will be deployed in July. 
These latter four lines are planned to be connected in September 2007. The detector is expected to be complete early 2008.

The line motions are monitored by acoustic devices (high frequency long base line LBL) and by inclinometers
 regularly spread along the line, allowing redundancy. The system allows a location of each OM with a precision close 
to 10~cm. Timing calibration is ensured by a network of laser and LED beacons~\cite{calibration}. According to 
the design specifications, a precision mesurement of $0.4$~ns is achieved which guaranties an angular resolution 
within expectations ($<0.5^{\rm o}$).

\begin{figure}[t]
\begin{center}
\includegraphics [width=0.4\textwidth]{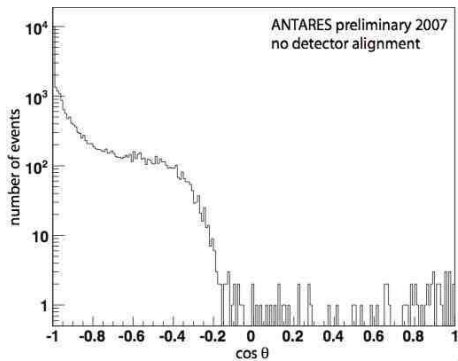}
\end{center}
\caption{The zenith angle distribution of data taken during Feb-May 2007 with a quality cut 
based on the fit likelihood. This preliminary reconstruction is based on the nominal positions of the OMs.
 Alignement data, now available, will considerably improve the recontruction efficiency.
While most of the tracks are reconstructed in the 
downward-going direction there is a steep fall around ${\rm cos}\theta=-0.2$ as expected from the flux 
of cosmic ray muons. Some upward going events are seen which are candidates for neutrino events.}
\label{zenith}
\end{figure}

\begin{figure*}[ht]
\begin{center}
\includegraphics [width=0.8\textwidth]{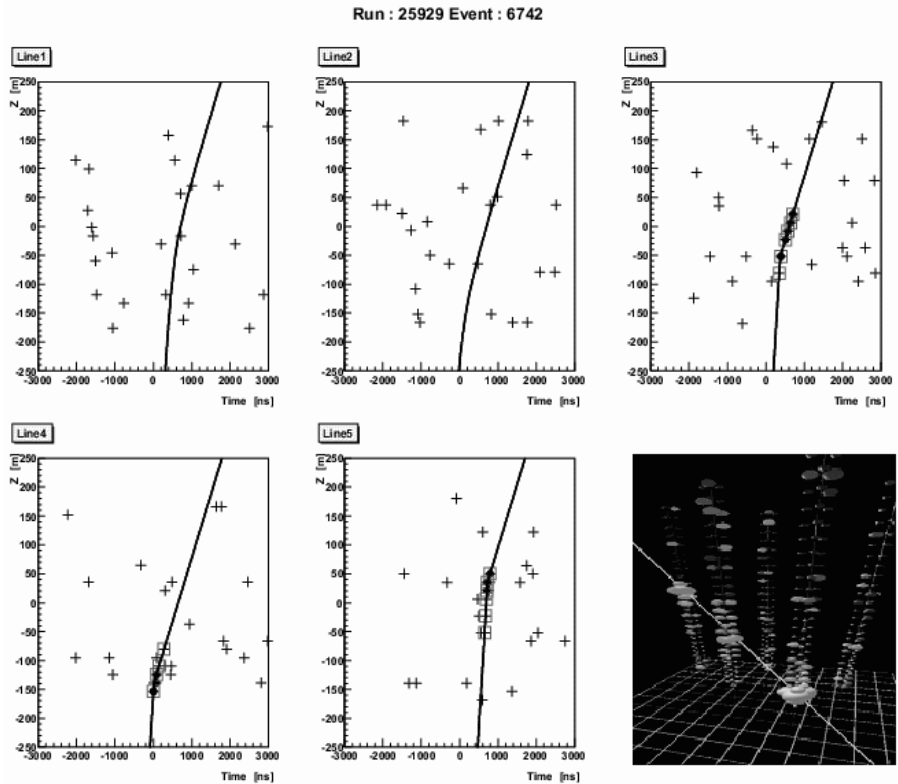}
\end{center}
\caption{Example of atmospheric neutrino induced muon candidate obtained with the 5 line detector. Each plot shows a single line hit distribution as a function of time. The bottom-right drawing is a 3D display of the same event. The muon trajectory is reconstructed upgoing with a zenith angle $35.4^{\rm o}$ away from vertical.}
\label{display}
\end{figure*}

The existing 5 line data are dominated by downward going muon bundles, 
the present trigger rate being roughly $1$~Hz. The reconstruction program fits a single track to these events under
 the assumption that light is emitted under the Cherenkov angle w.r.t the muon path. 
The angular distribution obtained, after quality cuts, is shown in figure~\ref{zenith}. As one can see, upward 
candidates are also present in the reconstructed sample. One of these neutrino candidates is displayed in figure~\ref{display}.

\section{Conclusions}

Great achievements have been made by the Antares collaboration in the last year. The detector is working in nominal mode 
with 5 lines and should be complete early 2008. Upward neutrino candidates have been found that validate the conceptual method and
 the chosen techniques. Very exciting times have started with a detector looking for neutrinos in a region of the celestial
 sky which has never been studied with such a level of sensitivity.

\bibliography{libros}
\bibliographystyle{unsrt}

\end{document}